\documentclass[pra,amsfonts,amssymb,amsmath,showpacs]{revtex4}

\begin{document}

\title{The two-dimensional hydrogen atom revisited}

\author{D.G.W.~Parfitt} \affiliation{School of Physics, University of
Exeter, Stocker Road, Exeter EX4 4QL, United Kingdom}

\author{M.E.~Portnoi}
\email[Electronic mail: ]{m.e.portnoi@ex.ac.uk}
\altaffiliation[Also at ]{A.F.~Ioffe
Physico-Technical Institute, St.~Petersburg, Russia.}
\affiliation{School of Physics, University of Exeter, Stocker Road,
Exeter EX4 4QL, United Kingdom}



\begin{abstract}
The bound state energy eigenvalues for the two-dimensional Kepler
problem are found to be degenerate. This ``accidental'' degeneracy is
due to the existence of a two-dimensional analogue of the
quantum-mechanical Runge-Lenz vector.  Reformulating the problem in
momentum space leads to an integral form of the Schr\"{o}dinger
equation.  This equation is solved by projecting the two-dimensional
momentum space onto the surface of a three-dimensional sphere. The
eigenfunctions are then expanded in terms of spherical harmonics, and
this leads to an integral relation in terms of special functions which
has not previously been tabulated. The dynamical symmetry of the
problem is also considered, and it is shown that the two components of
the Runge-Lenz vector in real space correspond to the generators of
infinitesimal rotations about the respective coordinate axes in
momentum space.
\end{abstract}

\maketitle

\section{Introduction}

A semiconductor quantum well under illumination is a
quasi-two-dimensional system, in which photoexcited electrons and
holes are essentially confined to a plane.  The mutual Coulomb
interaction leads to electron-hole bound states known as excitons,
which are extremely important for the optical properties of the
quantum well. The relative in-plane motion of the electron and hole
can be described by a two-dimensional Schr\"{o}dinger equation for a
single particle with a reduced mass.  
This is a physical realization of the two-dimensional hydrogenic problem, 
which originated as a purely theoretical construction \cite{F52}.
An important similarity with the three-dimensional hydrogen 
atom is the ``accidental'' degeneracy of the bound state energy levels. 
This degeneracy is due to the existence of the quantum-mechanical
Runge-Lenz vector, first introduced by Pauli \cite{P26} in three
dimensions, and indicates the presence of a dynamical symmetry of the
system.

The most important study relating to the hidden symmetry of the
hydrogen atom was that by Fock in 1935 \cite{F35}. He considered the
Schr\"{o}dinger equation in momentum space, which led to an integral
equation. Considering negative energy (bound-state) solutions, he
projected the three-dimensional momentum space onto the surface
of a four-dimensional hypersphere. 
After a suitable transformation of the wavefunction, the
resulting integral equation was seen to be invariant under
rotations in four-dimensional momentum space. Fock deduced that the
dynamical symmetry of the hydrogen atom is described by the
four-dimensional rotation group SO(4), which contains the geometrical
symmetry SO(3) as a subgroup. He related this hidden symmetry to the
observed degeneracy of the energy eigenvalues.

Shortly afterwards, Bargmann \cite{B36} made the connection between
Pauli's quantum mechanical Runge-Lenz vector and Fock's discovery of
invariance under rotations in four-dimensional momentum space. Fock's method
was also extended by Alliluev \cite{A58} to the case of $d$ dimensions
$(d\geqslant 2)$.  A comprehensive review concerning the symmetry of
the hydrogen atom was later given by Bander and Itzykson
\cite{BI66a,BI66b}, including a detailed group theoretical treatment
and extension to scattering states.

Improvements in semiconductor growth techniques over the subsequent
decades, which enabled the manufacture of effectively two-dimensional
structures, led to a resurgence of interest in the two-dimensional
hydrogen atom. The Runge-Lenz vector for this case was defined for the
first time \cite{YLC91}, and real-space solutions of the
Schr\"{o}dinger equation were applied to problems of atomic physics in
two-dimensions \cite{Y91}.

Recent studies have focused on diverse aspects of the hydrogenic
problem. The $d$-dimensional case has been reconsidered, leading to a
generalized Runge-Lenz vector (see \cite{A97} and references
therein). The algebraic basis of the dynamical symmetry has also been
given a thorough mathematical treatment \cite{D93,D98}.

In the present work we return to the two-dimensional problem, and use
the method of Fock to obtain a new integral relation in terms of
special functions.  The dynamical symmetry of the system is also
considered, and a new interpretation of the two-dimensional Runge-Lenz
vector is presented.

\section{Problem formulation}\label{S:sec2}

\subsection{Preliminaries}

The relative in-plane motion of an electron and hole, with effective
masses $m_e$ and $m_h$, respectively, may be treated as that of a
single particle with reduced mass $\mu=m_em_h/(m_e+m_h)$ and energy
$E$, moving in a Coulomb potential $V(\rho)$. The wavefunction of the
particle satisfies the stationary Schr\"{o}dinger equation
\begin{equation}\label{wannier}
\hat{H}\Psi(\boldsymbol{\rho})=\left[-\frac{1}{\rho}\frac{\partial}
{\partial\rho}\left(\rho\frac{\partial}{\partial\rho}\right)-\frac{1}
{\rho^2}\frac{\partial^2}{\partial\phi^2}+V(\rho)\right]
\Psi(\boldsymbol{\rho})=E\Psi(\boldsymbol{\rho}),
\end{equation}
where $(\rho,\phi)$ are plane polar coordinates.  Note that excitonic
Rydberg units are used throughout this paper, which leads to a
potential of the form $V(\rho)=-2/\rho$.

The eigenfunctions of Eq.~\eqref{wannier} are derived in Appendix
\ref{S:radial}. It is well known that the bound state energy levels
are of the form \cite{F52}:
\begin{equation}\label{elevel}
E=-\frac{1}{\left(n+1/2\right)^2},\qquad n=0,1,2,\ldots
\end{equation}
where $n$ is the principal quantum number. Notably, Eq.~\eqref{elevel} does
not contain explicitly the azimuthal quantum number $m$, which enters
the radial equation (see Appendix \ref{S:radial},
Eq.~\eqref{radialeqn}). Each energy level is $(2n+1)$-fold 
degenerate, the so-called accidental degeneracy.

It is convenient to introduce a vector operator corresponding to the
$z$-projection of the angular momentum,
$\hat{\mathbf{L}}_z=\mathbf{e}_z\hat{L}_z$, where $\mathbf{e}_z$ is a
unit vector normal to the plane of motion of the electron and hole. We
now introduce the two-dimensional analogue of the quantum-mechanical
Runge-Lenz vector as the dimensionless operator
\begin{equation}\label{dimless}
\hat{\mathbf{A}}=(\hat{\mathbf{q}}\times\hat{\mathbf{L}}_z-
\hat{\mathbf{L}}_z\times\hat{\mathbf{q}})-\frac{2}{\rho}\boldsymbol{\rho},
\end{equation}
where $\hat{\mathbf{q}}=-i\nabla$ is the momentum operator. Note that
$\hat{\mathbf{A}}$ lies in the plane and has Cartesian components
$\hat{A}_x$ and $\hat{A}_y$.

$\hat{L}_z$, $\hat{A}_x$ and $\hat{A}_y$ represent conserved
quantities and therefore commute with the Hamiltonian:
\begin{equation}
[\,\hat{H},\,\hat{L}_z\,]=[\,\hat{H},\,\hat{A}_x\,]=[\,\hat{H},\,\hat{A}_y\,]=0.
\end{equation}
They also satisfy the following commutation relations:
\begin{align}
[\,\hat{L}_z,\,\hat{A}_x\,]&=i\hat{A}_y, \\
[\,\hat{L}_z,\,\hat{A}_y\,]&=-i\hat{A}_x, \\
[\,\hat{A}_x,\,\hat{A}_y\,]&=-4i\hat{L}_z\hat{H}.
\end{align}

\subsection{Derivation of energy eigenvalues from $\hat{\mathbf{A}}$}

The existence of the non-commuting operators $\hat{A}_x$ and
$\hat{A}_y$, representing conserved physical quantities, implies that
the Runge-Lenz vector is related to the accidental degeneracy of the
energy levels in two dimensions \cite{LL77}. We now present a simple
interpretation of the hidden symmetry underlying this degeneracy.

For eigenfunctions of the Hamiltonian we can replace $\hat{H}$ by the
energy $E$, and defining
\begin{equation}
\hat{\mathbf{A}}'=\frac{\hat{\mathbf{A}}}{2\sqrt{-E}},
\end{equation}
we obtain the new commutation relations:
\begin{align}
[\,\hat{L}_z,\,\hat{A}_x'\,]&=i\hat{A}_y', \\
[\,\hat{L}_z,\,\hat{A}_y'\,]&=-i\hat{A}_x', \\
[\,\hat{A}_x',\,\hat{A}_y'\,]&=i\hat{L}_z.
\end{align}

If we now construct a three-dimensional vector operator
\begin{equation}
\hat{\mathbf{J}}=\hat{\mathbf{A}}'+\hat{\mathbf{L}}_z,
\end{equation}
then the components of $\hat{\mathbf{J}}$ satisfy the commutation
rules of ordinary angular momentum:
\begin{equation}
[\,\hat{J}_j,\,\hat{J}_k\,]=i\epsilon_{jkl}\hat{J}_l,
\end{equation}
where $\epsilon_{jkl}$ is the Levi-Civita symbol.

Noting that
$\hat{\mathbf{A}}'\cdot\hat{\mathbf{L}}_z=\hat{\mathbf{L}}_z\cdot\hat{\mathbf{A}}'=0$,
we have
\begin{equation}\label{jsq}
\hat{\mathbf{J}}^2={(\hat{\mathbf{A}}'+\hat{\mathbf{L}}_z)}{}^2={\hat{\mathbf{A}}
'}{}^2+\hat{\mathbf{L}}_z^2,
\end{equation}
where the operator $\hat{\mathbf{J}}^2$ has eigenvalues $j(j+1)$ and
commutes with the Hamiltonian.

We now make use of a special expression relating $\hat{\mathbf{A}}^2$
and $\hat{\mathbf{L}}_z^2$, the derivation of which is given in
Appendix \ref{S:asquared}:
\begin{equation}\label{asqeqn}
\hat{\mathbf{A}}^2=\hat{H}(4\hat{\mathbf{L}}_z^2+1)+4.
\end{equation}
Substituting in Eq.~\eqref{jsq} and again replacing $\hat{H}$ with
$E$, we obtain
\begin{equation}\label{jsqu}
\hat{\mathbf{J}}^2=-\frac{1}{4E}\left[E(4\hat{\mathbf{L}}_z^2+1)+4\right]+
\hat{\mathbf{L}}_z^2.
\end{equation}
Because $[\,\hat{H},\,\hat{\mathbf{J}}^2\,]=0$, an eigenfunction of
the Hamiltonian will also be an eigenfunction of $\hat{\mathbf{J}}^2$.
Operating with both sides of Eq.~\eqref{jsqu} on an eigenfunction of
the Hamiltonian, we obtain for the eigenvalues of $\hat{\mathbf{J}}^2$:
\begin{equation}
j(j+1)=-\left(\frac{1}{4}+\frac{1}{E}\right).
\end{equation}
Rearranging, and identifying $j$ with the principal quantum number
$n$, we obtain the correct expression for the energy eigenvalues:
\begin{equation}
E=-\frac{1}{\left(n+1/2\right)^2},\qquad n=0,1,2,\ldots
\end{equation}

Note that the $z$-component of $\hat{\mathbf{J}}$ is simply
$\hat{L}_z$. Recalling that the eigenvalues of $\hat{L}_z$ are denoted by $m$,
there are $(2j+1)$ values of $m$ for a given $j$. However, as $j=n$,
we see that there are $(2n+1)$ values of $m$ for a given energy, which
corresponds to the observed $(2n+1)$-fold degeneracy.

\section{Fock's method in two dimensions}\label{S:sec3}

\subsection{Stereographic projection}\label{SS:Fock}

The method of Fock \cite{F35}, in which a three-dimensional momentum
space is projected onto the surface of a four-dimensional hypersphere,
may be applied to our two-dimensional problem. We begin by defining a
pair of two-dimensional Fourier transforms between real space and
momentum space:
\begin{align}
\Phi(\mathbf{q})&=\int\Psi(\boldsymbol{\rho})e^{i\mathbf{q}\cdot
\boldsymbol{\rho}}\,d\boldsymbol{\rho},
\label{f1}\\
\Psi(\boldsymbol{\rho})&=\frac{1}{(2\pi)^2}\int\Phi(\mathbf{q})e^{-i\mathbf{q}
\cdot\boldsymbol{\rho}}\,d\mathbf{q}.
\label{f2}
\end{align}
We shall restrict our interest to bound states, and hence the energy
$E=-q_0^2$ will be negative.

Substitution of Eq.~\eqref{f2} in Eq.~\eqref{wannier} yields the
following integral equation for $\Phi(\mathbf{q})$:
\begin{equation}\label{intSE}
\left(q^2+q_0^2\right)\Phi(\mathbf{q})=\frac{1}{\pi}
\int\frac{\Phi(\mathbf{q'})\,d\mathbf{q'}}{|\mathbf{q}-\mathbf{q'}|}.
\end{equation}

The two-dimensional momentum space is now projected onto the surface
of a three-dimensional unit sphere centered at the origin, and so it is
natural to scale the in-plane momentum by $q_0$. Each point on a unit
sphere is completely defined by two polar angles, $\theta$ and $\phi$,
and the Cartesian coordinates of a point on the unit sphere are given
by
\begin{align}
u_x&=\sin\theta\cos\phi=\frac{2q_0q_x}{q^2+{q_0}^2},\label{c1} \\
u_y&=\sin\theta\sin\phi=\frac{2q_0q_y}{q^2+{q_0}^2},\label{c2} \\
u_z&=\cos\theta=\frac{q^2-{q_0}^2}{q^2+{q_0}^2}.\label{c3}
\end{align}

An element of surface area on the unit sphere is given by
\begin{equation}
d\Omega=\sin\theta\,d\theta\,d\phi=\left(\frac{2q_0}{q^2+{q_0}^2}\right)^2
\,d\mathbf{q},
\end{equation}
and the distance between two points transforms as:
\begin{equation}
|\mathbf{u}-\mathbf{u'}|=\frac{2q_0}{{(q^2+{q_0}^2)}^{1/2}
{(q^{\prime 2}+{q_0}^2)}^{1/2}}|\mathbf{q}-\mathbf{q'}|.
\end{equation}

If the wavefunction on the sphere is expressed as
\begin{equation}\label{trans}
\chi(\mathbf{u})=\frac{1}{\sqrt{q_0}}\left(\frac{q^2+{q_0}^2}
{2q_0}\right)^{3/2}\Phi(\mathbf{q}),
\end{equation}
then Eq.~\eqref{intSE} reduces to the simple form:
\begin{equation}\label{int}
\chi(\mathbf{u})=\frac{1}{2\pi
q_0}\int\frac{\chi(\mathbf{u'})\,d\Omega'}{|\mathbf{u}-\mathbf{u'}|}.
\end{equation}

\subsection{Expansion in spherical harmonics}

Any function on a sphere can be expressed in terms of spherical
harmonics, so for $\chi(\mathbf{u})$ we have
\begin{equation}\label{sph1}
\chi(\mathbf{u})=\sum_{l=0}^{\infty}\sum_{m=-l}^{l}A_{lm}Y_l^m(\theta,\phi),
\end{equation}
where $Y_l^m(\theta,\phi)$ are basically defined as in \cite{MW70}:
\begin{equation}
Y_{l}^{m}(\theta,\phi)=c_{lm}\sqrt{\frac{2l+1}{4\pi}\frac{(l-|m|)!}
{(l+|m|)!}}P_l^{|m|}(\cos\theta)e^{im\phi},
\end{equation}
where $P_n^{|m|}(\cos\theta)$ is an associated Legendre function as
defined in \cite{GR00}. The constant $c_{lm}$ is an arbitrary ``phase
factor". As long as $|c_{lm}|^2=1$ we are free to choose $c_{lm}$, and
for reasons which will become clear we set
\begin{equation}\label{phase}
c_{lm}=(-i)^{|m|}.
\end{equation}

The kernel of the integral in Eq.~\eqref{int} can also be
expanded in this basis as \cite{A85}:
\begin{equation}\label{sph2}
\frac{1}{|\mathbf{u}-\mathbf{u'}|}=\sum_{\lambda=0}^{\infty}
\sum_{\mu=-\lambda}^{\lambda}\frac{4\pi}{2\lambda+1}
Y_\lambda^\mu(\theta,\phi)Y_\lambda^{\mu*}(\theta ',\phi ').
\end{equation}

Substituting Eqs.~\eqref{sph1} and \eqref{sph2} into Eq.~\eqref{int}
we have
\begin{multline}\label{long}
\sum_{l=0}^{\infty}\sum_{m=-l}^{l}A_{lm}Y_l^m(\theta,\phi) \\
=\frac{2}{q_0}\sum_{l_1=0}^{\infty}\sum_{l_2=0}^{\infty}\sum_{m_1=-l_1}^{l_1}
\sum_{m_2=-l_2}^{l_2}\int\frac{1}{2l_2+1}A_{l_1m_1}Y_{l_1}^{m_1}(\theta',\phi')
Y_{l_2}^{m_2}(\theta,\phi)Y_{l_2}^{m_2*}(\theta',\phi')\,d\Omega'.
\end{multline}

We now make use of the orthogonality property of spherical harmonics
to reduce Eq.~\eqref{long} to the following:
\begin{equation}\label{short}
\sum_{l=0}^{\infty}\sum_{m=-l}^{l}A_{lm}Y_l^m(\theta,\phi)=\frac{2}{q_0}
\sum_{l_1=0}^{\infty}\sum_{m_1=-l_1}^{l_1}\frac{1}{2l_1+1}A_{l_1m_1}
Y_{l_1}^{m_1}(\theta,\phi).
\end{equation}
Multiplying both sides of Eq.~\eqref{short} by
$Y_n^{m'*}(\theta,\phi)$ and integrating over $d\Omega$ gives
\begin{equation}
A_{nm'}=\frac{2}{q_0(2n+1)}A_{nm'},
\end{equation}
where we have again used the orthogonality relation for spherical
harmonics. The final step is to rearrange for $q_0$ and identify the
index $n$ with the principal quantum number. This enables us to find
an expression for the energy in excitonic Rydbergs:
\begin{equation}
E=-q_0^2=-\frac{1}{\left(n+1/2\right)^2},\qquad n=0,1,2,\ldots
\end{equation}
This is seen to be identical to Eq. \eqref{elevel}.

For a particular value of $n$, the general solution of Eq.~\eqref{int}
can be expressed as
\begin{equation}\label{chin}
\chi_n(\mathbf{u})=\sum_{m=-n}^{n}A_{nm}Y_{n}^{m}(\theta,\phi).
\end{equation}
Each of the functions entering the sum in Eq.~\eqref{chin}
satisfies Eq.~\eqref{int} separately.
So, for each value of $n$ we have $(2n+1)$ linearly independent
solutions, and this explains the observed $(2n+1)$-fold degeneracy.

We are free to choose any linear combination of spherical harmonics for our
eigenfunctions, but for convenience we simply choose
\begin{equation}\label{eig}
\chi_{nm}(\mathbf{u})=A_{nm}Y_{n}^{m}(\theta,\phi).
\end{equation}
If we also require our eigenfunctions to be normalized as follows:
\begin{equation}
\frac{1}{(2\pi)^2}\int|\chi(\mathbf{u})|^2\,d\Omega=\frac{1}{(2\pi)^2}
\int\frac{q^2+q_0^2}{2q_0^2}|\Phi(\mathbf{q})|^2\,d\mathbf{q}=
\int|\Psi(\boldsymbol{\rho})|^2\,d\boldsymbol{\rho}=1,
\end{equation}
then Eq.~\eqref{eig} reduces to
\begin{equation}
\chi_{nm}(\mathbf{u})=2\pi Y_{n}^{m}(\theta,\phi).
\end{equation}

Applying the transformation in Eq.~\eqref{trans}, we can obtain an
explicit expression for the orthonormal eigenfunctions of Eq.~\eqref{intSE}:
\begin{equation}\label{phiq}
\Phi_{nm}(\mathbf{q})=c_{nm}\sqrt{2\pi\frac{(n-|m|)!}{(n+|m|)!}}
\left(\frac{2q_0}{q^2+q_0^2}\right)^{3/2}P_n^{|m|}(\cos\theta)e^{im\phi},
\end{equation}
where we have used the fact that $q_0=(n+1/2)^{-1}$, and $\theta$ and 
$\phi$ are defined by Eqs.~\eqref{c1}--\eqref{c3}.

\subsection{New integral relations}\label{SS:ints}

To obtain the real-space eigenfunctions $\Psi(\boldsymbol{\rho})$ we make
an inverse Fourier transform:
\begin{equation}\label{phitrans}
\Psi(\boldsymbol{\rho})=\frac{1}{(2\pi)^2}\int\Phi(\mathbf{q})
e^{-i\mathbf{q}\cdot\boldsymbol{\rho}}\,d\mathbf{q}=\frac{1}{(2\pi)^2}
\int_0^{2\pi}\int_0^{\infty}\Phi(\mathbf{q})e^{-iq\rho\cos\phi'}
\,q\,dq\,d\phi',
\end{equation}
where $\phi'$ is the azimuthal angle between the vectors
$\boldsymbol{\rho}$ and $\mathbf{q}$. However, if we now substitute
Eq.~\eqref{phiq} into this expression we have to be careful with our
notation. The angle labeled $\phi$ in Eq.~\eqref{phiq} is actually
related to $\phi'$ via
\begin{equation}\label{phirel}
\phi=\phi'+\phi_\rho,
\end{equation}
where $\phi_\rho$ is the azimuthal angle of the vector
$\boldsymbol{\rho}$, which can be treated as constant for the purposes
of our integration.

Taking this into account, the substitution of Eq.~\eqref{phiq} into
Eq.~\eqref{phitrans} yields:
\begin{equation}\label{phi2}
\Psi(\boldsymbol{\rho})=\frac{c_{nm}}{(2\pi)^{3/2}}
\sqrt{\frac{(n-|m|)!}{(n+|m|)!}}\,e^{im\phi_\rho}\int_0^{2\pi}
\int_0^{\infty}\left(\frac{2q_0}{q^2+q_0^2}\right)^{3/2}
P_n^{|m|}(\cos\theta)e^{i(m\phi'-q\rho\cos\phi')}\,q\,dq\,d\phi'.
\end{equation}

From Eq.~\eqref{c3} we obtain
\begin{equation}\label{theta}
P_n^{|m|}(\cos\theta)=P_n^{|m|}\left(\frac{q^2-{q_0}^2}{q^2+{q_0}^2}\right),
\end{equation}
and we use the following form of Bessel's integral \cite{A85}:
\begin{equation}\label{bes}
\int_0^{2\pi}e^{i(m\phi'-q\rho\cos\phi')}\,d\phi'=2\pi(-i)^m
J_m(q\rho),
\end{equation}
where $J_m(q\rho)$ is a Bessel function of the first kind of order
$m$.  Substituting Eqs.~\eqref{theta} and \eqref{bes} into
Eq.~\eqref{phi2} leads to
\begin{equation}\label{rho1}
\Psi(\boldsymbol{\rho})=\frac{c_{nm}(-i)^m}{\sqrt{2\pi}}
\sqrt{\frac{(n-|m|)!}{(n+|m|)!}}\,e^{im\phi_\rho}
\int_0^{\infty}\left(\frac{2q_0}{q^2+q_0^2}\right)^{3/2}P_n^{|m|}
\left(\frac{q^2-{q_0}^2}{q^2+{q_0}^2}\right)J_m(q\rho)\,q\,dq.
\end{equation}

We now make a change of variables, $x=q_0\rho$ and $y=q^2/q_0^2$, so
that Eq.~\eqref{rho1} becomes
\begin{equation}\label{rad}
\Psi(\boldsymbol{\rho})=c_{nm}(-1)^{n+m}(-i)^m\sqrt{\frac{q_0(n-|m|)!}
{\pi(n+|m|)!}}\,e^{im\phi_\rho}\int_0^{\infty}P_n^{|m|}
\left(\frac{1-y}{1+y}\right)\frac{J_m(x\sqrt{y})}{(1+y)^{3/2}}\,dy,
\end{equation}
where we have used the fact that \cite{A85}:
\begin{equation}
P_n^{|m|}\left(\frac{y-1}{y+1}\right)=(-1)^{n+m}P_n^{|m|}
\left(\frac{1-y}{1+y}\right).
\end{equation}

If we now equate the expression for $\Psi(\boldsymbol{\rho})$ in 
Eq.~\eqref{rad} with that derived in Appendix \ref{S:radial}, we obtain
the following:
\begin{equation}\label{newint1}
c_{nm}(-1)^{n+m}(-i)^m\int_0^{\infty}P_n^{|m|}\left(\frac{1-y}{1+y}\right)
\frac{J_m(x\sqrt{y})}{(1+y)^{3/2}}\,dy=
\frac{(2x)^{|m|}e^{-x}}{n+1/2}L_{n-|m|}^{2|m|}(2x).
\end{equation}
The value of $c_{nm}$ chosen earlier in Eq.~\eqref{phase} ensures that
both sides of Eq.~\eqref{newint1} are numerically equal. If we
restrict our interest to $m\geqslant 0$ then the relation simplifies
to
\begin{equation}\label{newint2}
\int_0^{\infty}P_n^{m}\left(\frac{1-y}{1+y}\right)\frac{J_m(x\sqrt{y}
)}{(1+y)^{3/2}}\,dy=\frac{(-1)^n(2x)^{m}e^{-x}}{n+1/2}L_{n-m}^{2m}(2x),
\qquad n,m=0,1,2,\ldots;\:m\leqslant n.
\end{equation}

As far as we can ascertain, this integral relation between special
functions has not previously been tabulated. For $n,m=0$ we recover
the known integral relation \cite{GR00}:
\begin{equation}
\int_0^{\infty}\frac{J_0(x\sqrt{y})}{(1+y)^{3/2}}\,dy=2e^{-x}.
\end{equation}

\section{Dynamical symmetry}\label{S:sec4}

\subsection{Infinitesimal generators}

Consider now a vector $\mathbf{u}$ from the origin to a point on the
three-dimensional unit sphere defined in Sec.~\ref{SS:Fock}. If this
vector is rotated through an infinitesimal angle $\alpha$ in the
$(u_xu_z)$ plane, we have a new vector
\begin{equation}\label{defn}
\mathbf{u}'=\mathbf{u}+\mathbf{\delta u},
\end{equation}
where the components of $\mathbf{u}$ are given in
Eqs.~\eqref{c1}--\eqref{c3}, and
\begin{equation}
\delta\mathbf{u}=\alpha\,\mathbf{e}_y\times\mathbf{u}.
\end{equation}
This rotation on the sphere corresponds to a change in the two-dimensional
momentum from $\mathbf{q}$ to $\mathbf{q'}$.
The Cartesian components of Eq.~\eqref{defn} are then found to be
\begin{align}
u_{x}^{\prime}=\frac{2q_0 q_x^{\prime}}{q'{}^2+q_0^2}&=
\frac{2q_0 q_x}{q^2+q_0^2}+\alpha\frac{q^2-q_0^2}{q^2+q_0^2}, \\ 
u_{y}^{\prime}=\frac{2q_0 q_y^{\prime}}{q'{}^2+q_0^2}&=
\frac{2q_0 q_y}{q^2+q_0^2}, \\
u_{z}^{\prime}=\frac{q'{}^2-q_0^2}{q'{}^2+q_0^2}&=
\frac{q^2-q_0^2}{q^2+q_0^2}-\alpha\frac{2q_0 q_x}{q^2+q_0^2},
\end{align}
where $q^2=q_x^2+q_y^2$.

After some manipulation we can also find the components of
$\delta\mathbf{q}=\mathbf{q}'-\mathbf{q}$:
\begin{align}\label{deltas}
\delta q_x&=\alpha\frac{q^2-q_0^2-2q_x^2}{2q_0}, \\
\delta q_y&=-\alpha\frac{q_x q_y}{q_0}.
\end{align}

The corresponding change in $\Phi(\mathbf{q})$ is given by
\begin{equation}
\delta\Phi(\mathbf{q})=\frac{\alpha}{(q^2+q_0^2)^{3/2}}
\left(\frac{q^2-q_0^2-2q_x^2}{2q_0}\frac{\partial}{\partial
q_x}-\frac{q_x q_y}{q_0}\frac{\partial}{\partial
q_y}\right)\left[(q^2+q_0^2)^{3/2}\Phi(\mathbf{q})\right].
\end{equation}
We can write this as
\begin{equation}
\delta\Phi(\mathbf{q})=-\frac{i}{2q_0}\alpha\hat{\mathcal{A}}_x
\Phi(\mathbf{q}),
\end{equation}
where the infinitesimal generator is given by
\begin{equation}
\hat{\mathcal{A}}_x=\frac{i}{(q^2+q_0^2)^{3/2}}\left((q^2-q_0^2-2q_x^2)
\frac{\partial}{\partial
q_x}-2q_xq_y\frac{\partial}{\partial q_y}\right)(q^2+q_0^2)^{3/2}.
\end{equation}

We now make use of the following operator expression in the momentum 
representation:
\begin{equation}
\hat{\boldsymbol{\rho}}=\mathbf{e}_x\hat{x}+\mathbf{e}_y\hat{y}=
i\nabla_q,
\end{equation}
and the commutation relation
\begin{equation}
[\,\hat{\boldsymbol{\rho}},\,f(\mathbf{q})\,]=i\nabla_q f,
\end{equation}
to derive a more compact expression for $\hat{\mathcal{A}}_x$:
\begin{equation}
\hat{\mathcal{A}}_x=(q^2-q_0^2)\hat{x}-2q_x(\mathbf{q}\cdot
\hat{\boldsymbol{\rho}})-3iq_x.
\end{equation}
By considering an infinitesimal rotation in the $(u_yu_z)$ plane we
can obtain a similar expression for $\hat{\mathcal{A}}_y$:
\begin{equation}
\hat{\mathcal{A}}_y=(q^2-q_0^2)\hat{y}-2q_y(\mathbf{q}\cdot
\hat{\boldsymbol{\rho}})-3iq_y.
\end{equation}

These expressions operate on a particular energy eigenfunction with
eigenvalue $-q_0^2$. If we move the constant $-q_0^2$ to the right and
replace it with the Hamiltonian in momentum space, $\hat{\mathcal{H}}$:
\begin{align}
\hat{\mathcal{A}}_x&=q^2\hat{x}+\hat{x}\hat{\mathcal{H}}-2q_x(\mathbf{q}
\cdot\hat{\boldsymbol{\rho}})-3iq_x,
\label{ax} \\
\hat{\mathcal{A}}_y&=q^2\hat{y}+\hat{y}\hat{\mathcal{H}}-2q_y(\mathbf{q}
\cdot\hat{\boldsymbol{\rho}})-3iq_y,
\label{ay}
\end{align}
then $\hat{\mathcal{A}}_x$ and $\hat{\mathcal{A}}_y$ can operate on
any linear combination of eigenfunctions.

\subsection{Relation to Runge-Lenz vector}

Recall the definition of the two-dimensional Runge-Lenz vector in real space:
\begin{equation}\label{rl2}
\hat{\mathbf{A}}=(\hat{\mathbf{q}}\times\hat{\mathbf{L}}_z-
\hat{\mathbf{L}}_z\times\hat{\mathbf{q}})-\frac{2}{\rho}\boldsymbol{\rho}.
\end{equation}
Using $\hat{\mathbf{L}}_z=\boldsymbol{\rho}\times\hat{\mathbf{q}}$,
and the following identity for the triple product of three vectors:
\begin{equation}
\mathbf{a}\times(\mathbf{b}\times\mathbf{c})=(\mathbf{a}
\cdot\mathbf{c})\mathbf{b}-(\mathbf{a}\cdot\mathbf{b})\mathbf{c},
\end{equation}
we can apply the commutation relation
$[\,\boldsymbol{\rho},\,\hat{\mathbf{q}}\,]=i$ to rewrite Eq.~\eqref{rl2}
in the form:
\begin{equation}\label{rl3}
\hat{\mathbf{A}}=\hat{\mathbf{q}}^2\boldsymbol{\rho}+\boldsymbol{\rho}
\left(\hat{\mathbf{q}}^2-\frac{2}{\rho}\right)-2\hat{\mathbf{q}}
(\hat{\mathbf{q}}\cdot\boldsymbol{\rho})-3i\hat{\mathbf{q}}.
\end{equation}

If we now return to the expression for the real-space Hamiltonian in 
Eq.~\eqref{wannier}, it is apparent that we may substitute
\begin{equation}\label{ham2}
\hat{\mathbf{q}}^2-2/\rho=\hat{H}
\end{equation}
in Eq.~\eqref{rl3} to yield
\begin{equation}
\hat{\mathbf{A}}=\hat{\mathbf{q}}^2\boldsymbol{\rho}+\boldsymbol{\rho}
\hat{H}-2\hat{\mathbf{q}}(\hat{\mathbf{q}}\cdot\boldsymbol{\rho})-3i
\hat{\mathbf{q}}.
\end{equation}
Comparing this with Eqs.~\eqref{ax} and \eqref{ay}, it is evident that
the two components of the Runge-Lenz vector in real space correspond
to the generators of infinitesimal rotations in the $(u_xu_z)$ and
$(u_yu_z)$ planes.

\section{Conclusion}\label{S:sec5}

We have shown that the accidental degeneracy in the energy eigenvalues
of the two-dimensional Kepler problem may be explained by the
existence of a planar analogue of the familiar three-dimensional
Runge-Lenz vector.  By moving into momentum space and making a
stereographic projection onto a three-dimensional sphere, 
a new integral relation in terms of special functions has 
been obtained, which to our knowledge has not previously been
tabulated. We have also demonstrated explicitly that 
the components of the two-dimensional Runge-Lenz vector in real space 
are intimately related to infinitesimal rotations in three-dimensional
momentum space.

\appendix

\section{Solution of real-space Schr\"{o}dinger equation}\label{S:radial}

We apply the method of separation of variables to Eq.~\eqref{wannier},
making the substitution
\begin{equation}
\Psi(\boldsymbol{\rho})=R(\rho)\Phi(\phi).
\end{equation}
Introducing a separation constant $m^2$, we can obtain the angular
equation
\begin{equation}
\frac{d^2\Phi}{d\phi^2}+m^2\Phi=0,
\end{equation}
with the solution
\begin{equation}
\Phi(\phi)=\frac{1}{\sqrt{2\pi}}e^{im\phi}.
\end{equation}

The corresponding radial equation (with $E=-q_0^2$) is
\begin{equation}\label{radialeqn}
\frac{d^2
R}{d\rho^2}+\frac{1}{\rho}\frac{dR}{d\rho}+\left(\frac{2}{\rho}-q_0^2-
\frac{m^2}{\rho^2}\right)R=0.
\end{equation}
We make the substitution
\begin{equation}
R(\rho)=C\rho^{|m|}e^{-q_{0}\rho}w(\rho),
\end{equation}
where $C$ is a normalization constant. This leads to the equation
\begin{equation}
\rho\frac{d^2w}{d\rho^2}+(2|m|+1-2q_0\rho)\frac{dw}{d\rho}+(2-2|m|q_0-q_0)w=0.
\end{equation}
Making a final change of variables $\beta=2q_0\rho$, we obtain
\begin{equation}\label{hyper}
\beta\frac{d^2w}{d\beta^2}+(2|m|+1-\beta)\frac{dw}{d\beta}+
\left(\frac{1}{q_0}-|m|-\frac{1}{2}\right)w=0.
\end{equation}
This is the confluent hypergeometric equation \cite{GR00}, which has 
two linearly independent solutions. If we choose the solution which is 
regular at the origin, then this becomes a polynomial of finite degree if
$q_0=(n+1/2)^{-1}$ with $n=0,1,2,\ldots$  Eq.~\eqref{hyper} then
becomes the associated Laguerre equation \cite{A85}, 
the solutions of which are the associated Laguerre polynomials:
\begin{equation}
w=L_{n-|m|}^{2|m|}(\beta)=L_{n-|m|}^{2|m|}(2q_0\rho).
\end{equation}

We can now write the real-space wavefunction in the form
\begin{equation}
\Psi_{nm}(\boldsymbol{\rho})=\frac{C}{2\pi}\rho^{|m|}e^{-q_0\rho}
L_{n-|m|}^{2|m|}(2q_0\rho)e^{im\phi_\rho},
\end{equation}
where the reason for the subscript on $\phi$ is explained in
Sec.~\ref{SS:ints}.

To normalize this wavefunction we need to make use of the integral
\cite{A85}:
\begin{equation}
\int_0^{\infty}e^{-2q_0\rho}(2q_0\rho)^{2|m|+1}L_{n-|m|}^{2|m|}(2q_0\rho)
L_{n-|m|}^{2|m|}(2q_0\rho)\,d(2q_0\rho)=\frac{(n+|m|)!}{(n-|m|)!}(2n+1).
\end{equation}
The normalized wavefunctions are therefore:
\begin{equation}
\Psi_{nm}(\boldsymbol{\rho})=\sqrt{\frac{q_0^3(n-|m|)!)}{\pi(n+|m|)!}}
(2q_0\rho)^{|m|}e^{-q_0\rho}L_{n-|m|}^{2|m|}(2q_0\rho)e^{im\phi_\rho},
\end{equation}
satisfying the following orthogonality condition:
\begin{equation}
\int\Psi_{n_1m_1}^*(\boldsymbol{\rho})\Psi_{n_2m_2}(\boldsymbol{\rho})\,d
\boldsymbol{\rho}=\delta_{n_1n_2}\delta_{m_1m_2}.
\end{equation}

\section{Derivation of Eq.~\ref{asqeqn}}\label{S:asquared}

From Eq.~\eqref{dimless} we have
\begin{align}\label{asq}
\hat{\mathbf{A}}^2&=\left[(\hat{\mathbf{q}}\times\hat{\mathbf{L}}_z-
\hat{\mathbf{L}}_z\times\hat{\mathbf{q}})-\frac{2}{\rho}
\boldsymbol{\rho}\right]^2 \\
&=[2(\hat{\mathbf{q}}\times\hat{\mathbf{L}}_z)-i\hat{\mathbf{q}}]^2-
\frac{2}{\rho}\boldsymbol{\rho}\cdot[2(\hat{\mathbf{q}}
\times\hat{\mathbf{L}}_z)-i\hat{\mathbf{q}}]-\frac{2}{\rho}[2(\hat{\mathbf{q}}
\times\hat{\mathbf{L}}_z)-i\hat{\mathbf{q}}]\cdot\boldsymbol{\rho}+4.
\notag
\end{align}
We further expand as follows:
\begin{align}\label{exp1}
[2(\hat{\mathbf{q}}\times\hat{\mathbf{L}}_z)-i\hat{\mathbf{q}}]^2&=
4(\hat{\mathbf{q}}\times\hat{\mathbf{L}}_z)^2-2i\hat{\mathbf{q}}
\cdot(\hat{\mathbf{q}}\times\hat{\mathbf{L}}_z)-2i(\hat{\mathbf{q}}
\times\hat{\mathbf{L}}_z)\cdot\hat{\mathbf{q}}-\hat{\mathbf{q}}^2 \\
&=4\hat{\mathbf{q}}^2\hat{\mathbf{L}}_z^2+2\hat{\mathbf{q}}^2-\hat{\mathbf{q}}^2
=\hat{\mathbf{q}}^2(4\hat{\mathbf{L}}_z^2+1),
\notag
\end{align}
and
\begin{equation}\label{exp2}
-\frac{2}{\rho}\boldsymbol{\rho}\cdot[2(\hat{\mathbf{q}}\times
\hat{\mathbf{L}}_z)-i\hat{\mathbf{q}}]-\frac{2}{\rho}[2(\hat{\mathbf{q}}
\times\hat{\mathbf{L}}_z)-i\hat{\mathbf{q}}]\cdot\boldsymbol{\rho}
=-\frac{2}{\rho}(4\hat{\mathbf{L}}_z^2+1).
\end{equation}
Substituting Eqs.~\eqref{exp1} and \eqref{exp2} into Eq.~\eqref{asq} gives
\begin{equation}
\hat{\mathbf{A}}^2=\hat{\mathbf{q}}^2(4\hat{\mathbf{L}}_z^2+1)-
\frac{2}{\rho}(4\hat{\mathbf{L}}_z^2+1)+4,
\end{equation}
which, from Eq.~\eqref{ham2}, is just
\begin{equation}
\hat{\mathbf{A}}^2=\hat{H}(4\hat{\mathbf{L}}_z^2+1)+4.
\end{equation}


\bibliographystyle{apsrev} \bibliography{kepler2d}

\end{document}